\documentstyle[amssymb,aps,12pt]{revtex}

\begin{document}
\author{T. P. Cheng$^{*}$ and Ling-Fong Li$^{\dagger }$}
\address{$^{*}${\small Department of Physics and Astronomy, University of Missouri,
St. Louis, MO 63121}\\
$^{\dagger }${\small Department of Physics, Carnegie Mellon University,
Pittsburgh, PA 15213}}
\title{Chiral Quark Model of Nucleon Spin-Flavor Structure\\
with SU(3) and Axial-U(1) Breakings \bigskip }
\date{CMU-HEP97-01, hep-ph/9701xxx}
\maketitle

\begin{abstract}
\bigskip \bigskip

{\bf Abstract: }The chiral quark model with a nonet of Goldstone bosons can
yield an adequate description of the observed proton flavor and spin
structure. In a previous publication we have compared the results of a SU(3)
symmetric calculation with the phenomenological findings based on
experimental measurements and SU(3) symmetry relations. In this paper we
discuss their SU(3) and axial U(1) breaking corrections. Our result
demonstrates the broad consistency of the chiral quark model with the
experimental observations of the proton spin-flavor structure. With two
parameters, we obtain a very satifactory fit to the $F/D$ ratios for the
octet baryon masses and for their axial vector couplings, as well as the
different quark flavor contributions to the proton spin. The result also can
account for, not only the light quark asymmetry $\bar{u}-\bar{d}$, but also
the strange quark content $\bar{s},$ of the proton sea. SU(3) breaking is
the key in reconciling the $\bar{s}$ value as measured in the neutrino charm
production and that as deduced from the pion nucleon sigma term$.$
\end{abstract}

\section{Introduction}

A significant part of the nucleon structure study involves non-perturbative
QCD. As the structure problem may be very complicated when viewed directly
in terms of the fundamental degrees of freedom (current quarks and gluons),
it may well be useful to break the problem into two stages. One first
identify the relevant effective degrees of freedom in terms of which the
description for such non-perturbative physics will be simple, intuitive and
phenomenologically correct; at the next stage, one then elucidates the
relations between these non-perturbative degrees of freedom in terms of the
QCD quarks and gluons. For the nonperturbative phenomena taking place just
inside the confinement scale, the chiral quark model ($\chi $QM) suggests
that the relevant degrees of freedom as being the internal Goldstone bosons
(GB), constituent quarks, which can be thought of as just the quarks
propagating in the QCD vacuum, for this energy range with its chiral
condensate. The hope is that, without waiting for a final explication of the
detailed mechanism for chiral symmetry breaking and confinement, we can yet
achieve a simple description of the hadron structure.

Our investigation has been built upon the prior work by Eichten, Hinchliffe
and Quigg\cite{EHQ}, who applied the $\chi $QM idea\cite{ManoGeorgi} to the
proton flavor and spin problem. In our previous publication\cite{CL95}, we
have argued on phenomenological and theoretical grounds for the inclusion of
a flavor-SU(3) singlet meson with a coupling to the constituent quark having
an opposite sign to the octet coupling, $g_{0}\simeq -g_{8}$. In this
picture, we have been able to account for much of the observed proton spin
and flavor structure which is puzzling from the view point of naive
constituent quark model: the $\overline{u}$ - $\overline{d}$ asymmetry (as
measured by the deviation from the Gottfried sum rule\cite{NMC}\cite
{Gottfried} and by the cross section difference of the Drell-Yan processes
on proton and neutron targets\cite{NA51}), a significant strange quark
content $\overline{s}$ (as indicated \cite{Cheng76} by the value of the
pion-nucleon sigma term $\sigma _{\pi N}$ \cite{sfactor}\cite{sigma45MeV}),
as well as the various quark flavor contributions to the proton spin (as
deduced from the violation of the Ellis-Jaffe sum rule\cite{EJsum} -\cite
{EllisK}). Furthermore, the chiral quark model predicts that the antiquarks
in a nucleon are not significantly polarized. We have suggested that this
feature is consistent with our picture of the baryon magnetic moments being
built up from those of the constituent quarks having Dirac moments\cite{CL96}%
. In the meantime, SMC have presented their data on their semi-inclusive
spin asymmetry measurements indicating the antiquark polarizations $\Delta 
\overline{u}$ and $\Delta \overline{d}$ being consistent with zero\cite
{SMC96}, thus providing further support for this $\chi $QM explanation of
the proton spin-flavor puzzle.

The phenomenological success of this chiral quark model requires that the
basic interactions between Goldstone bosons and constituent quarks being
feeble enough that the perturbative description is applicable. This is so,
even though the underlying phenomena of spontaneous chiral symmetry breaking
and confinement are, obviously, non-perturbative.

Our previous calculation has been performed in the SU(3) symmetric limit,
and we have compared the results to phenomenological values which have been
deduced by using SU(3) symmetry relations as well. For example the various
quark flavor contributions to the proton spin, such as the strange quark
polarization $\Delta s,$ have been extracted after using the SU(3) symmetric
F/D ratio for hyperon decays\cite{EJsum}. Similarly, the extraction of the
strange quark content $f_{s}=\left( s+\overline{s}\right) /\left[
\sum_{q}\left( q+\overline{q}\right) \right] $ from the experimental value
of $\sigma _{\pi N}$ involves the same sort of SU(3) symmetry relation among
octet baryon masses\cite{Cheng76}. It is gratifying that the agreements are
in the 20\% to 30\% range, indicating that the broken-U(3) chiral quark
picture\cite{CL95} is, perhaps, on the right track.

To take the next step is, however, much more difficult. The phenomenological
values $\Delta s$ and $f_{s}$ are sensitive to SU(3) breaking effects, which
can only be introduced in the extraction process in a model-dependent way.
Consequently these phenomenological quantities would have large uncertainty
if no SU(3) symmetry is assumed. Correspondingly, it is difficult to perform
a $\chi $QM calculation away from the SU(3) symmetric limit: SU(3) breaking
is introduced by different quark masses $m_{s}>m_{u,d}$ and by the
non-degenerate Goldstone boson masses $M_{K,\eta }>M_{\pi }$, and the axial
U(1) breaking by $M_{\eta ^{\prime }}>M_{K,\eta }.$ Since these are
Goldstone modes propagating inside hadrons, they are expected to have
effective masses different from the physical pseudoscalar meson masses.
Apparently, in order to study such symmetry breaking effect in detail, one
would need a theory of these GBs propagating in the intermediate range
between the confinement scale and the energy scale below which the
spontaneous chiral symmetry breaking takes place: $\Lambda _{conf}<\Lambda
<\Lambda _{\chi sb}.$

Nevertheless, some effort has already been made in the study of the symmetry
breaking effects on the phenomenological values. Several authors have
obtained results suggesting that both $\Delta s$ and $f_{s}$ will be reduced
by such effects\cite{su3br} -\cite{Gass81}. It is then worthwhile to see
what sort of pattern would the chiral quark model suggest for such
corrections, to see whether they are compatible with the modified physical
data, as well as yielding an overall agreement with phenomenology at the
better-than-20\% level. Our purpose in this paper is to present such a
schematic SU(3) and axial U(1) breaking calculation to demonstrate the broad
consistency of our chiral quark model with the observational data.

\section{Chiral QM calculation with SU(3) breaking}

The SU(3) breaking effects will be introduced\cite{song} in the amplitudes
for GB emission by a quark, simply through the insertion of a suppression
factor: $\epsilon $ for kaons, $\delta $ for eta, and $\zeta $ for eta prime
mesons, as these strange-quark-bearing GBs are presumably more massive than
the pions. Thus the probability $a\propto $ $\left| g_{8}\right| ^{2}$are
modifies for processes involving strange quarks, as shown in Table 1, where
we have already substituted-in the quark content of the GBs. The suppression
factors enter into the amplitudes for $u_{+}\rightarrow \left( u\overline{u}%
\right) _{0}u_{-}$ and $u_{+}\rightarrow \left( d\overline{d}\right)
_{0}u_{-}$ processes, {\em etc.} because they also receive contributions
from the $\eta $ and $\eta ^{\prime }$ GBs.

\subsection{The flavor content}

From Table 1, one can immediately read off the antiquark number $\overline{q}
$ in the proton after the emission of one GB by the initial proton state $%
\left[ \left( 2u+d\right) \rightarrow ....\right] :\;$ 
\begin{eqnarray}
\overline{u} &=&\frac{1}{12}\left[ \left( 2\zeta +\delta +1\right)
^{2}+20\right] a,  \label{ubar} \\
\overline{d} &=&\frac{1}{12}\left[ \left( 2\zeta +\delta -1\right)
^{2}+32\right] a,  \label{dbar} \\
\overline{s} &=&\frac{1}{3}\left[ \left( \zeta -\delta \right)
^{2}+9\epsilon ^{2}\right] a.  \label{sbar}
\end{eqnarray}
For the quark number in the proton, we have 
\begin{equation}
u=2+\overline{u},\;\;\;\;\;d=1+\bar{d},\;\;\;\;\;s=\bar{s},  \label{flvnum}
\end{equation}
because, in the quark sea, the quark and antiquark numbers of a given flavor
are equal. We shall also make use the notion ``quark flavor fraction in a
proton''\ $f_{q}\;$defined as 
\begin{equation}
f_{q}=\frac{\left\langle \bar{q}q\right\rangle _{p}}{\left\langle \bar{u}u+%
\bar{d}d+\bar{s}s\right\rangle _{p}}=\frac{q+\bar{q}}{3+2\left( \bar{u}+\bar{%
d}+\bar{s}\right) },  \label{fq}
\end{equation}
where $q^{\prime }s$ in the proton matrix elements $\left\langle \bar{q}%
q\right\rangle _{p}\;$are the quark field operators, and in the last term
they stand for the quark numbers in the proton.

\subsection{The spin content}

In the limit when interactions are negligible, we have the proton wave
function for the spin-up state as 
\[
\left| p_{+}\right\rangle =\frac{1}{\sqrt{6}}\left( 2\left|
u_{+}u_{+}d_{-}\right\rangle -\left| u_{+}u_{-}d_{+}\right\rangle -\left|
u_{-}u_{+}d_{+}\right\rangle \right) 
\]
This implies that the probability of finding $u_{+},\;u_{-},\;d_{+},$ and$%
\;d_{-}$ are $\frac{5}{3},\;\frac{1}{3},\;\frac{1}{3},\;$and $\frac{2}{3},$
respectively, leading to the naive quark model prediction of $\Delta
u=u_{+}-\;u_{-}=\frac{4}{3},$ $\Delta d=-\frac{1}{3}$ and $\Delta s=0.$
After emission of one GB, which flips the quark helicity (see Table 1), we
have 
\begin{eqnarray}
\Delta u &=&\frac{4}{3}\left[ 1-\Sigma P_{1}\right] +\frac{1}{3}P_{1}\left(
u_{-}\rightarrow u_{+}\right) +\frac{2}{3}P_{1}\left( d_{-}\rightarrow
u_{+}\right) -\frac{5}{3}P_{1}\left( u_{+}\rightarrow u_{-}\right) -\frac{1}{%
3}P_{1}\left( d_{+}\rightarrow u_{-}\right)  \nonumber \\
\Delta d &=&-\frac{1}{3}\left[ 1-\Sigma P_{1}\right] +\frac{1}{3}P_{1}\left(
u_{-}\rightarrow d_{+}\right) +\frac{2}{3}P_{1}\left( d_{-}\rightarrow
d_{+}\right) -\frac{5}{3}P_{1}\left( u_{+}\rightarrow d_{-}\right) -\frac{1}{%
3}P_{1}\left( d_{+}\rightarrow d_{-}\right)  \nonumber \\
\Delta s &=&\frac{1}{3}P_{1}\left( u_{-}\rightarrow s_{+}\right) +\frac{2}{3}%
P_{1}\left( d_{-}\rightarrow s_{+}\right) -\frac{5}{3}P_{1}\left(
u_{+}\rightarrow s_{-}\right) -\frac{1}{3}P_{1}\left( d_{+}\rightarrow
s_{-}\right)  \label{delcal}
\end{eqnarray}
where $P_{1}\left( d_{+}\rightarrow s_{-}\right) =\epsilon ^{2}a$ is the
probability of a spin-up $d$ quark flipping into a spin-down $s$ quark
(through the emission of $K^{+}$), as displayed in Table 1, {\em etc}. The
combination $\left[ 1-\Sigma P_{1}\right] $ stands for the probability of
``no GB emission'', with $\Sigma P_{1}$ being the total probability of
emitting one GB ($\pi ^{+},\;\pi ^{0},\;K,\;\eta ,\;\eta ^{\prime }$): 
\begin{equation}
1-\Sigma P_{1}=1-\left( 1+\frac{1}{2}+\epsilon ^{2}+\frac{\delta ^{2}}{6}+%
\frac{\zeta ^{2}}{3}\right) a.  \label{no-emi}
\end{equation}
After plugging in the probabilities in Eq.(\ref{delcal}), we obtain the
various quark contributions to the proton spin: 
\begin{eqnarray}
\Delta u &=&\frac{4}{3}-\frac{21+4\delta ^{2}+8\zeta ^{2}+12\epsilon ^{2}}{9}%
a  \label{delu} \\
\Delta d &=&-\frac{1}{3}-\frac{6-\delta ^{2}-2\zeta ^{2}-3\epsilon ^{2}}{9}a
\label{deld} \\
\Delta s &=&-\epsilon ^{2}a  \label{dels}
\end{eqnarray}

\subsection{The SU(3) parameters: D \& F}

It has been pointed out in our previous paper\cite{CL95} that, since a SU(3)
symmetric calculation would not alter the relative strength of quantities
belonging to the same SU(3) multiplet, our symmetric calculation cannot be
expected to improve on the naive quark model results such as the axial
vector coupling ratio $F/D=2/3,$ which differs significantly from the
generally quoted phenomenological value of $F/D=0.575\pm 0.016$\cite
{FtoDaxial}$.$ To account for this difference we must include the SU(3)
breaking terms: 
\begin{eqnarray}
\frac{F}{D} &=&\frac{\Delta u-\Delta s}{\Delta u+\Delta s-2\Delta d} 
\nonumber \\
&=&\frac{2}{3}\cdot \frac{6-a\left( 2\delta ^{2}+4\zeta ^{2}+\frac{1}{2}%
\left( 3\epsilon ^{2}+21\right) \right) }{6-a\left( 2\delta ^{2}+4\zeta
^{2}+9\epsilon ^{2}+3\right) }.  \label{ftod}
\end{eqnarray}
Similarly discussion holds for the $F/D$ ratio for the octet baryon masses.
Here we choose to express this in terms of the quark flavor fractions as
defined by Eq.(\ref{fq}): 
\begin{eqnarray}
\frac{f_{3}}{f_{8}} &=&\frac{f_{u}-f_{d}}{f_{u}+f_{d}-2f_{s}}=\frac{%
1+2\left( \bar{u}-\bar{d}\right) }{3+2\left( \bar{u}+\bar{d}-2\bar{s}\right) 
}  \nonumber \\
&=&\frac{1}{3}\cdot \frac{3+2a\left[ 2\zeta +\delta -3\right] }{3+2a\left[
2\zeta \delta +\frac{1}{2}\left( 9-\delta ^{2}-12\epsilon ^{2}\right)
\right] }.  \label{fratio}
\end{eqnarray}
The SU(6) prediction $\frac{1}{3}$ should be compared to the
phenomenological value of $0.21$\cite{ftodbmass}.

In the SU(3) symmetry limit of $\delta =\epsilon =1,$ we can easily check
that Eqs.(\ref{ftod}) and (\ref{fratio}) reduce, independent of $a$ and $%
\zeta $, to their naive quark model values$.$

\section{Numerics}

What impact do these SU(3) and U(1)$_{A}$ breaking suppression factors have
on the comparison of chiral quark results with phenomenological quantities?
Here we shall put in a few numbers. Again our purpose is not so much as
finding the precise best-fit values, but using some simple choice of
parameters to illustrate the structure of chiral quark model. In this spirit
we shall pick the suppression factors for the $K$ and $\eta $ amplitudes to
be comparable: $\epsilon \simeq \delta $. As for the suppression factor $%
\zeta $ for the $\eta ^{\prime }$ emission amplitude, since the symmetric
calculation\cite{CL95} favors $\zeta \simeq -1$, and since $\eta ^{\prime }$
is extra heavy, {\em i.e.} axial U(1) is broken, we will simply pick $\zeta
\simeq -\frac{1}{2}\epsilon .$ Thus, for the numerical consideration, we
start with the simple approximation of 
\begin{equation}
\epsilon =\delta =-2\zeta .  \label{apx}
\end{equation}

Perhaps the most significant part of the chiral quark picture is its
explanation of the isospin asymmetry of the quark sea, which the NMC has
measured to be\cite{NMC} 
\begin{equation}
\bar{u}-\bar{d}=\frac{3}{2}\left( \int_{0}^{1}dxF_{2}^{p-n}\left( x\right) -%
\frac{1}{3}\right) \simeq -0.15.  \label{udbardiff}
\end{equation}
From Eqs.(\ref{ubar}) - (\ref{dbar}), the $\chi $QM expression for this
difference is 
\begin{equation}
\bar{u}-\bar{d}=\left[ \frac{2\zeta +\delta }{3}-1\right] a.  \label{ud-diff}
\end{equation}
With the approximation of Eq.(\ref{apx}), this suggests that we pick the
emission probability $a\simeq 0.15.$ As for the suppression factors, we
shall take the illustrative value of $\epsilon =\delta =-2\zeta \simeq 0.6.$
If one wishes to, one can interpret these values as the relative strength of
the propagator factors: 
\[
\Gamma _{\pi }:\Gamma _{K}:\Gamma _{\eta }:\Gamma _{\eta ^{\prime
}}=1:\epsilon :\delta :\left| \zeta \right| 
\]
where 
\[
\Gamma _{\pi }=\frac{1}{\left\langle Q^{2}\right\rangle +M_{\pi }^{2}}%
,\;etc.\;\;\;\;\;\;\text{with \ }\left\langle Q^{2}\right\rangle \simeq
0.35\,GeV^{2}. 
\]

In Table 2, we summarize the results of such a numerical calculation. They
are compared to the phenomenological values, and to the predictions by the
naive quark model and by the $\chi $QM with SU(3) symmetry, respectively.

We should mention that, in this crude model calculation, we cannot specified
the detailed Bjorken-$x$ dependence of the various quark densities. Namely,
all the densities should be taken as those averaged over the entire range of 
$x$. In this connection, one should be careful in making a comparison of the
antiquark density ratio of $\bar{u}/\bar{d}$ , which our model (with the
stated parameters) yields a value of $0.63,$ while the NA51 Collaboration%
\cite{NA51} measured it to have a value of $0.51\pm 0.04\pm 0.05$ at a
specific point of $x=0.15.$

\section{Discussion \& Conclusion}

In our previous publication\cite{CL95}, we have demonstrated that the chiral
quark model with a nonet of GBs can, in the SU(3) symmetric limit with the
singlet coupling $g_{0}\simeq -g_{8}$, yield an adequate accounting of the
observed proton spin and flavor structure. In this paper, we have presented
a calculation which takes into account, schematically, the SU(3) symmetry
breaking effects due to the heavier strange quark, $m_{s}>m_{u,d}$ and $%
M_{K,\eta }>M_{\pi },$ as well as the axial U(1) breaking due to $M_{\eta
^{\prime }}>M_{K,\eta }$. We find the resulting phenomenology having been
significantly improved.

\subsection{F/D ratios}

We wish to emphasize that the calculation presented here is more than just
an exercise in parametrizing the experimental data. After fixing the two
constants by the measured values, we have been able to reproduce several
other phenomenological quantities. Our point is that the broken-U(3) $\chi $%
QM with $m_{s}>m_{u,d}$ has just the right structure to account for the
overall pattern of the experimental data. For example, it has been clear
that in this model the SU(3) breaking terms are needed to account for the
deviation $F/D$ ratios from the SU(6) predictions\cite{CL95}. But there is
no {\em a priori }reason to expect the correction to either increase or
decrease the ratio. However, our schematic calculations shows that this
model have the right structure to make the correction in just the right
direction. Consider the axial vector coupling $F/D$ ratio of Eq.(\ref{ftod}%
). To simplify our presentation, let us expand it in powers of the emission
probability $a:$%
\[
\frac{F}{D}=\frac{2}{3}\left[ 1-\frac{5}{4}\left( 1-\epsilon ^{2}\right)
a+O\left( a^{2}\right) \right] . 
\]
The desired correction is for the above $\left[ ...\right] $ factor to be
less than one, see Table 2. This is precisely what the $\chi $QM with $%
m_{s}>m_{u,d}$ would lead one to expect because of the inequality $\epsilon
^{2}<1.$ Similar statement can also be made for the ratio in Eq.(\ref{fratio}%
): 
\[
\frac{f_{3}}{f_{8}}=\frac{1}{3}\left\{ 1-\frac{1}{3}\left[ \left( 1-\delta
^{2}\right) +2\left( 1-2\zeta \right) \left( 1-\delta \right) +12\left(
1-\epsilon ^{2}\right) \right] a+O\left( a^{2}\right) \right\} . 
\]

Parenthetically, the axial vector coupling $3F-D=\Delta u+\Delta d-2\Delta
s\equiv \Delta _{8}$ has the structure $\Delta _{8}=\Delta _{8}^{\left(
0\right) }+\Delta _{8}^{\left( 1\right) }$ with the symmetric term $\Delta
_{8}^{\left( 0\right) }=1-\frac{2\zeta ^{2}+7}{3}a$ and the SU(3) breaking
correction being 
\[
\Delta _{8}^{\left( 1\right) }=\frac{1}{3}\left[ \left( 1-\delta ^{2}\right)
-3\left( 1-\epsilon ^{2}\right) \right] a\simeq -\frac{2}{3}\left(
1-\epsilon ^{2}\right) a<0. 
\]
Namely, in our $\chi $QM, $\Delta _{8}$ is reduced by SU(3) breaking
effects. This is again compatible the trend found for the phenomenological
extracted value - although our model indicates that this reduction is rather
moderate (from a symmetric value of $0.67$ to corrected value of $0.57$,
approximately) rather than the $50\%$ reduction as suggested in one of the $%
1/N_{c}$ studies\cite{ucsd}.

\subsection{Strange quark content \& polarization}

The $\chi $QM naturally suggests that the nucleon strange quark content $%
\bar{s}$ and polarization $\Delta s$ magnitude are lowered by the SU(3)
breaking effects as they are directly proportional to the amplitude
suppression factors, see Eqs.(\ref{sbar}) and (\ref{dels}). This is just the
trend found in the extracted phenomenological values. Gasser\cite{Gass81},
for instance, using a chiral loop model to calculate the SU(3) breaking
correction to the Gell-Mann-Okubo baryon mass formula, finds that the
no-strange-quark limit-value of $\left( \sigma _{\pi N}\right) _{0}$ is
modified from $25\,MeV$ to $35\,MeV$; this reduces $f_{s}$ from $0.18$ to $%
0.10,$ for a phenomenological value of $\sigma _{\pi N}=45\,MeV$\cite
{sfactor}\cite{sigma45MeV}. It matches closely our numerical calculation
with the illustrative parameters, see Table 2.

The strange quark content can also be expressed as the relative abundance of
the strange to non-strange quarks in the sea, which in this model is given
as 
\begin{equation}
\lambda \equiv \frac{\bar{s}}{\frac{1}{2}\left( \bar{u}+\bar{d}\right) }=4%
\frac{\left( \zeta -\delta \right) ^{2}+9\epsilon ^{2}}{\left( 2\zeta
+\delta \right) ^{2}+27}\simeq 1.6\epsilon ^{2}=0.6.  \label{lambda-s}
\end{equation}
This can be compared to the strange quark content as measured by the CCFR
Collaboration in their neutrino charm production experiment\cite{sbarexpt} 
\begin{equation}
\kappa \equiv \frac{\left\langle x\bar{s}\right\rangle }{\frac{1}{2}\left(
\left\langle x\bar{u}\right\rangle +\left\langle x\bar{d}\right\rangle
\right) }=0.477\pm 0.063,\;\;\;\text{where\ \ \ }\left\langle x\bar{q}%
\right\rangle =\int_{0}^{1}x\bar{q}\left( x\right) dx,  \label{kappa-s}
\end{equation}
which is often used in the global QCD reconstruction of parton distributions%
\cite{MRS}. The same experiment found no significant difference in the
shapes of the strange and non-strange quark distributions\cite{sbarexpt}: 
\[
x\bar{s}\left( x\right) \propto \left( 1-x\right) ^{\alpha }\frac{x\bar{u}%
\left( x\right) +x\bar{d}\left( x\right) }{2},
\]
with the shape parameter being consistent with zero, $\alpha =-0.02\pm 0.08.$
Thus, it is reasonable to use the CCFR findings to yield 
\begin{equation}
\lambda \simeq \kappa \simeq \frac{1}{2},  \label{s-content}
\end{equation}
which is a bit less than, but still compatible with, the value in Eq.(\ref
{lambda-s})\cite{soft-s}.

A number of authors have pointed out that phenomenologically extracted value
of strange quark polarization $\Delta s$ is sensitive to possible SU(3)
breaking corrections. While the effect is model-dependent, various
investigators\cite{su3br} -\cite{ucsd} all conclude that SU(3) breaking
correction tends to lower the magnitude of $\Delta s$. Some even suggested
the possibility of $\Delta s\simeq 0$ being consistent with experimental
data. Our calculation indicates that, while $\Delta s$ may be smaller than $%
0.10$, it is not likely to be significantly smaller than $0.05.$ To verify
this prediction, it is then important to pursue other phenomenological
methods that allow the extraction of $\Delta s$ without the need of SU(3)
relations. We recall that the elastic neutrino scatterings\cite{KapMano},
and the measurements of longitudinal polarization of $\Lambda $ in the
semi-inclusive process of $\bar{\nu}N\rightarrow \mu \Lambda +X$ \cite{WA58}
have already given support to a nonvanishing and negative $\Delta s.\,$Such
experimentation and phenomenological analysis should be pursued further\cite
{lambdabar}.

\subsection{Down quark polarization}

It is also interesting to examine the SU(3) breaking effect on the spin
contribution $\Delta d$, which should have only an indirect dependence on
the strange quark. Without SU(3) breaking, we have 
\[
\left( \Delta d\right) ^{\left( 0\right) }=-\frac{1}{3}-\frac{2}{9}\left(
1-\zeta ^{2}\right) a 
\]
which can hardly yield a $\Delta d$ value significantly more negative than $%
-1/3$ as required by phenomenology, whether in the simple $\chi $QM with an
octet of GBs ($\zeta =0$)$,$ or the broken-U(3) model with $\zeta =-1$. But
Eq.(\ref{deld}) clearly shows that it is the emission of
strange-quark-bearing mesons that contributes the ``wrong sign''. Hence, the
suppression of such emissions, when we take $m_{s}>m_{u,d}$ into account,
will make the $d$ quarks in the sea more negatively polarized, see Table 2.
Calculationally, the strange-quark-bearing mesons enter into the expression
for $\Delta d$ (with the wrong sign) through the probability factor for ``no
GB emission'' as given in (\ref{no-emi}).

\subsection{The role of SU(3)-singlet GB}

For the axial U(1) breaking, we made the parameter choice of $\zeta $ $%
\simeq -\frac{\epsilon }{2}\simeq -0.3.$ It implies that a satisfactory
phenomenology can be obtained with a strongly suppressed $\eta ^{\prime }$
amplitude. In what sense then are we required to extend the traditional $%
\chi $QM with an octet of GBs to the broken-U(3) version of the model? We
observe that if we set $\zeta =0$, namely a decoupled $\eta ^{\prime }$,
while the numerical results for $\Delta q$'s and $f_{s}$ remain quite
acceptable, the $\left( \bar{u}-\bar{d}\right) _{\zeta =0}=-0.12$ becomes
rather a poor fit to the known phenomenology$.$ Indeed we find it difficult
to get a good fit to all the phenomenological values with $\zeta =0:$ For
example, if we fix up the Gottfried sum rule violation with a some
adjustment of parameter: $a=0.175$ and $\epsilon =\delta \simeq \frac{1}{2},$
we then over-correct $f_{3}/f_{8}$ to $0.17,\,f_{s}$ to $0.06$ and $\kappa $
to $0.36,\;etc.$ (Generally speaking. it is the flavor, rather then the
spin, structure that is more sensitive to the $\zeta $ value.) Nevertheless,
it is difficult to justify the inclusion of the $\eta ^{\prime }$ meson
based on such crude numerical fit. We suggest that it is the overall
theoretical consistency that requires the inclusion of the SU(3)-singlet GB.
For example, from the view point of $1/N_{c}$ expansion, in the leading term
we have nine unmixed GBs. The next order nonplanar correction must be
included to break this U(3) symmetry --- and its attendant SU(3) symmetric
quark sea\cite{EHQ}\cite{CL95}, which is phenomenological undesirable ---
and to give the singlet an extra heavy mass (through the axial anomaly). In
our previous SU(3) symmetric calculation, we found that a choice of $%
g_{0}\simeq -g_{8}$ yield an adequate fit for a phenomenology derived at the
SU(3) symmetric level; in this paper a significantly better description has
been obtained after taking into account of SU(3) and U(1)$_{A}$ breakings.
All this shows that our broken-U(3) chiral quark model possesses a
consistent structure that can yield satisfactory phenomenological
descriptions at different levels of approximation.

\begin{center}
\medskip

{\bf ACKNOWLEDGMENTS}

\smallskip
\end{center}

This work is supported at CMU by the U. S. Department of Energy (Grant No.
DOE-ER/40682-127), at UM-St. Louis by an University of Missouri Research
Board award.\newpage

\begin{center}
$
\begin{tabular}{|c|c|c|}
\hline
$u_{+}\rightarrow $ & $d_{+}\rightarrow $ & Probability \\ \hline
$u_{+}\rightarrow \left( u\overline{d}\right) _{0}d_{-}$ & $d_{+}\rightarrow
\left( d\overline{u}\right) _{0}u_{-}$ & $a$ \\ 
$u_{+}\rightarrow \left( u\overline{s}\right) _{0}s_{-}$ & $d_{+}\rightarrow
\left( d\overline{s}\right) _{0}s_{-}$ & $\epsilon ^{2}a$ \\ 
$u_{+}\rightarrow \left( u\overline{u}\right) _{0}u_{-}$ & $d_{+}\rightarrow
\left( d\overline{d}\right) _{0}d_{-}$ & $\left( \frac{\delta +2\zeta +3}{6}%
\right) ^{2}a$ \\ 
$u_{+}\rightarrow \left( d\overline{d}\right) _{0}u_{-}$ & $d_{+}\rightarrow
\left( u\overline{u}\right) _{0}d_{-}$ & $\left( \frac{\delta +2\zeta -3}{6}%
\right) ^{2}a$ \\ 
$u_{+}\rightarrow \left( s\overline{s}\right) _{0}u_{-}$ & $d_{+}\rightarrow
\left( s\overline{s}\right) _{0}d_{-}$ & $\left( \frac{\delta -\zeta }{3}%
\right) ^{2}a$ \\ \hline
\end{tabular}
$

Table 1.

\bigskip

\begin{tabular}{|c|c|c|c|c|}
\hline
&  &  & $\chi $QM & $\chi $QM \\ 
& Phenomenological & Naive & SU$_{3}$ symmetric & broken SU$_{3}$ \\ 
& values & QM & $\epsilon =\delta =-\zeta =1$ & $\epsilon =\delta =-2\zeta
=0.6$ \\ 
&  &  & $a=0.11$ & $a=0.15$ \\ \hline
$\overline{u}-\overline{d}$ & $0.147\pm 0.026$ & $0\,$ & $0.146$ & $0.15$ \\ 
$\bar{u}/\bar{d}$ & $\left( 0.51\pm 0.09\right) _{x=0.15}$ & $1$ & $0.56$ & $%
0.63$ \\ 
$\frac{2\bar{s}}{\bar{u}+\bar{d}}$ & $\simeq 0.5$ & $0\,$ & $1.86$ & $0.60$
\\ 
$\sigma _{\pi N}:f_{s}$ & $0.18\pm 0.60\;\left( \downarrow ?\right) $ & $0$
& $0.19$ & $0.09$ \\ 
$f_{3}/f_{8}$ & $0.21\pm 0.05$ & $\frac{1}{3}$ & $\frac{1}{3}$ & $0.20$ \\ 
$g_{A}$ & $1.257\pm 0.03$ & $\frac{5}{3}$ & $1.12$ & $1.28$ \\ 
$\left( F/D\right) _{A}$ & $0.575\pm 0.016$ & $\frac{2}{3}$ & $\frac{2}{3}$
& $0.57$ \\ 
$\left( 3F-D\right) _{A}$ & $0.60\pm 0.07$ & $1$ & $0.67$ & $0.57$ \\ 
$\Delta u$ & $0.82\pm 0.06$ & $\frac{4}{3}$ & $0.78$ & $0.87$ \\ 
$\Delta d$ & $-0.44\pm 0.06$ & $-\frac{1}{3}$ & $-0.33$ & $-0.41$ \\ 
$\Delta s$ & $-0.11\pm 0.06\;\left( \downarrow ?\right) $ & $0$ & $-0.11$ & $%
-0.05$ \\ 
$\Delta \bar{u},\;\Delta \bar{d}$ & $-0.02\pm 0.11$ & $0$ & $0$ & $0$ \\ 
\hline
\end{tabular}

Table 2.

\newpage

{\bf TABLE CAPTION}
\end{center}

\begin{description}
\item[Table 1. ]  Transition probability for GB emission by constituent
quarks, with $a$ being that for the process $u_{+}\rightarrow \pi ^{+}d_{-},$
and with other processes reduced by SU(3) breaking suppression factors. The
subscripts $\pm $ represent the helicities of the quarks, being parallel or
anti-parallel to the proton helicity. The subscript $0$ indicates that the
quark and antiquark pair combine to form a spin zero state. Hence the
antiquarks, in the leading order of perturbation, have no net polarization.

\item[Table 2. ]  {\em \ }Comparison of $\chi $QM with phenomenological
values. Antiquark number difference $\overline{u}-\overline{d}\,$ follows
from the violation of Gottfried sum rule as measured by NMC\cite{NMC}. The $%
\chi $QM results for the antiquark density ratio $\bar{u}/\bar{d}$ are the $x
$-averaged quantities$,$ while the NA51 Collaboration\cite{NA51} measurement
is at a specific point of $x=0.15.$ The strange to nonstrange quark ratio in
the sea $\frac{2\bar{s}}{\bar{u}+\bar{d}}$ is from the CCFR measurement and
analysis\cite{sbarexpt} as discussed in the text, see Eq.(\ref{s-content}).
The strange quark fraction{\em \ }$f_{s}$ value is based on $\sigma _{\pi
N}=45\,MeV$ and the no-strange-quark limit-value of $\left( \sigma _{\pi
N}\right) _{0}=25\,MeV$, calculated by using the SU(3) symmetric baryon mass 
$F/D$ ratio\cite{sfactor}\cite{sigma45MeV}, and the quark-fraction ratio $%
f_{3}/f_{8}$ is similarly calculated by using the octet baryon masses\cite
{ftodbmass}. The axial vector coupling $F$ and $D$ are from Ref.\cite
{FtoDaxial}. Quark spin contribution $\Delta q$'s, based on the SU(3)
symmetric axial vector coupling $F/D$ ratio, are from summary review in \cite
{EllisK}. The antiquark polarization values $\Delta \bar{u}$ and $\Delta 
\bar{d}$ are from the recent SMC measurement on semi-inclusive processes\cite
{SMC96}. Possible downward revision of the phenomenological values by SU(3)
breaking effects, as discussed in the text, are indicated by the symbol $%
\left( \downarrow ?\right) $.
\end{description}


\begin{references}
\bibitem{sfactor}  The strange quark fraction in the proton is related to $%
\sigma _{\pi N}$ as $f_{s}=\frac{\sigma _{\pi N}-\left( \sigma _{\pi
N}\right) _{0}}{3\sigma _{\pi N}-\left( \sigma _{\pi N}\right) _{0}},$ where 
$\left( \sigma _{\pi N}\right) _{0}$ is the no-strange-quark limit value
related to the octet baryon masses and current quark mass ratio by $\left(
\sigma _{\pi N}\right) _{0}=\frac{3M_{8}}{1-r}\simeq 25\,MeV$ with $%
M_{8}=M_{\Lambda }-M_{\Xi }\simeq -200\,MeV$ and $r=m_{s}/m_{u,d}\simeq 25.$
\bibitem{song}  X. Song, in an University of Virginia Report (to be
published), has already initiated such a calculation, but with a suppression
factor only for kaon amplitudes. Namely, his calculation corresponds to ours
in the limit of $\delta =1$ and $\zeta =-1.2$. It seems to us, because the $%
\eta $ and $\eta ^{\prime }$ mesons (although carrying no net strangeness
quantum number) are more massive than the $K$ mesons, this is not a very
interesting limit to consider.
\bibitem{ftodbmass}  The octet quark-flavor fraction ratio is related to the
octet baryon masses: $\frac{f_{3}}{f_{8}}=\frac{\left\langle u\bar{u}-d\bar{d%
}\right\rangle }{\left\langle u\bar{u}+d\bar{d}-2s\bar{s}\right\rangle }=%
\frac{1}{3}\frac{2M_{N}+M_{\Sigma }-3M_{\Lambda }}{2M_{N}-M_{\Sigma
}-M_{\Lambda }}=0.21.$
\bibitem{lambdabar}  In this connection, it's also important to pursue
experimental measurements to check the $\chi $QM prediction for a vanishing
longitudinal polarization of $\bar{\Lambda}$ in the semi-inclusive processes
reflecting the proton spin property of $\Delta \bar{s}=0.$
\end{references}
\end{document}